\documentclass[reprint,superscriptaddress, nofootinbib,longbibliography]{revtex4-2}
\pdfoutput=1
\usepackage{graphicx}
\usepackage{amssymb}
\usepackage[linktoc=none]{hyperref}
\usepackage[dvipsnames]{xcolor}
\usepackage{xcolor}
\usepackage{dcolumn}
\usepackage[english]{babel}
\usepackage[utf8x]{inputenc}
\usepackage[T1]{fontenc}
\usepackage{lmodern}
\usepackage{eqnarray,amsmath,physics,bbold}
\usepackage{ dsfont }
\usepackage{appendix}
\usepackage{xcolor}
\usepackage{breqn}
\usepackage{comment}
\usepackage{physics}
\usepackage{relsize}
\usepackage{amsmath}
\usepackage{caption}
\usepackage{siunitx}
\usepackage{verbatim}
\usepackage{float}
\usepackage[labelformat=simple]{subcaption}

\DeclareCaptionLabelFormat{subcaptionlabel}{\normalfont(\textbf{#2}\normalfont)}
\usepackage{hyperref}
 \hypersetup{
 colorlinks = true, % Colours links instead of ugly boxes
 urlcolor = blue, % Colour for external hyperlinks
 linkcolor = blue, % Colour of internal links
 citecolor = red % Colour of citations
 }

\DeclareMathAlphabet{\mathpzc}{OT1}{pzc}{m}{it}

\newcommand{\kvec}{\mathbf{k}}

\newcommand{\INFN}{INFN, Sezione di Napoli, Gruppo Collegato di Salerno, I-84126 Napoli, Italy}

\newcommand{\UNISA}{Physics Department ``E.R. Caianiello'', University of Salerno, Via Giovanni Paolo II, 132, I-84084 Fisciano, SA, Italy}

\newcommand{\CNR}{CNR-SPIN, Via Giovanni Paolo II, 132, I-84084 Salerno, Italy }

\begin{document}
\title{Boltzmann theory of the inverse Edelstein effect in a two-dimensional Rashba gas
}

\author{Irene Gaiardoni}
\email{igaiardoni@unisa.it}
\affiliation{\UNISA}

\author{Mattia Trama}
\email{mtrama@unisa.it}
\affiliation{\UNISA}

\author{Alfonso Maiellaro}
\affiliation{\UNISA}
\affiliation{\CNR}

\author{Claudio Guarcello}
\affiliation{\UNISA}
\affiliation{\INFN}

\author{Francesco Romeo}
\affiliation{\UNISA}
\affiliation{\INFN}

\author{Roberta Citro}
%\email{rocitro@unisa.it}
\affiliation{\UNISA}
\affiliation{\CNR}
\affiliation{\INFN}

\begin{abstract}
We investigate the inverse Edelstein effect in a non-homogeneous system consisting of a ferromagnetic layer coupled to a Rashba two-dimensional electron gas. Within a semiclassical Boltzmann framework, we derive analytical expressions for the charge and spin currents and analyze their dependence on key parameters such as the chemical potential and the Rashba coupling strength. We show how interfacial exchange and spin–orbit interactions jointly control the efficiency of spin-to-charge conversion, leading to distinct regimes characterized by qualitatively different transport responses. A central outcome of our work is the availability of closed-form analytical results, which provide direct physical insight and enable a transparent and quantitative benchmarking with experiments on complex oxide interfaces, such as LaAlO$_3$/SrTiO$_3$.  

\end{abstract}

\maketitle

\section{Introduction}
In recent years, spin–charge interconversion phenomena have become a central topic in spintronics, as they enable the electrical generation, manipulation, and detection of spin currents \cite{doi:10.1126/science.1065389, PhysRevLett.61.2472,PhysRevB.39.4828,RevModPhys.76.323, 10.1093/acprof:oso/9780198568216.001.0001, soumyanarayanan2016emergent, sinova2004universal}. In this context, interfaces have emerged as particularly effective platforms, owing to the strong symmetry breaking and enhanced spin–orbit coupling. Among these systems, oxide-based interfaces have proven especially promising. For example, LaAlO$_3$/SrTiO$_3$ and LaAlO$_3$/KTaO$_3$ have garnered significant attention due to their unique electronic properties~\cite{gariglio2018spin, trama2022gate, trama2022tunable, trier2022oxide, 10.21468/SciPostPhys.17.4.101}.
At these interfaces, an electric field perpendicular to the plane breaks the inversion symmetry, leading to the formation of a two-dimensional electron gas (2DEG) with a Rashba spin-orbit coupling (RSOC)\cite{rashba1960spin}, which plays a crucial role into the spin–charge interconvertion phenomena~\cite{amin2016spin,vaz2019mapping, inoue2003diffuse,vicente2021spin,varotto2022direct,bibes2011ultrathin, gambardella2011current, seibold2017theory,2th8-nyz8,  trama2023effect, sxyb-5rcy}. Thus, the ability to manipulate the spin and charge in these systems opens up new avenues for spin-orbitronic applications~\cite{zhai2023large, guarcello2024probing, maiellaro2023hallmarks, 
bychkov1984properties, zpvy-t4d4, PhysRevB.110.184503, https://doi.org/10.1002/adma.202000818, 10.21468/SciPostPhys.17.4.101}. 
From a theoretical perspective, spin–charge conversion at interfaces has been extensively investigated in 2DEGs with RSOC, which provide a minimal and well-controlled platform to capture the essential microscopic mechanisms. In such systems, the lack of inversion symmetry leads to spin-split Fermi surfaces characterized by a chiral spin texture [see Fig.~\ref{Edelstein}(a)], whereby each Fermi contour carries a distinct spin polarization. This peculiar momentum–spin locking naturally enables spin–charge interconversion phenomena, most notably the direct and inverse Edelstein effects (DEE and IEE, respectively). The DEE, also referred to as the inverse galvanic effect, describes the generation of a non-equilibrium in-plane spin polarization in response to an applied electric field and has been the subject of extensive theoretical investigation~\cite{edelstein1990spin,leiva2023spin,aronov1989nuclear,johansson2021spin,johansson2016theoretical,gambardella2011current,trama2022tunable, sxyb-5rcy, PhysRevB.110.184503, gaiardoni2025edelstein}.
By Onsager reciprocity, its reciprocal counterpart, i.e, the IEE, corresponds to the conversion of an injected spin accumulation into a transverse charge current~\cite{shen2014microscopic,kato2004current,silov2004current,gorini2012onsager,sanchez2013spin,vignale2016theory}. Owing to its direct relevance for electrical spin detection and its sensitivity to the underlying spin texture, the IEE represents a particularly powerful probe of spin–orbit-coupled two-dimensional systems and constitutes the primary focus of the present work.
\begin{figure} [t]
\centering
\includegraphics[width=0.45\textwidth]{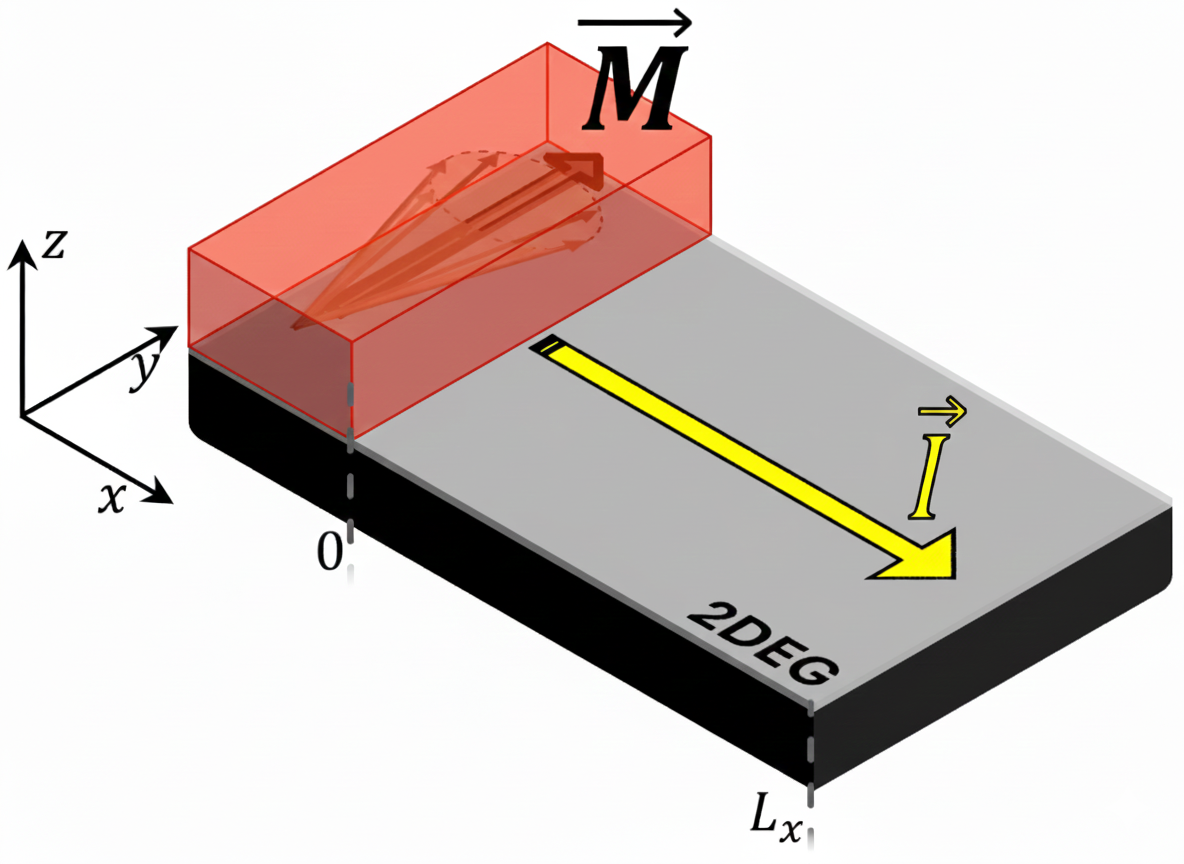}
 \caption{ 
Schematic of the setup used to study the inverse Edelstein effect (IEE). A ferromagnetic metal is placed in proximity to a two-dimensional electron gas (2DEG) with strong spin–orbit coupling. The magnetization of the ferromagnet undergoes precession around the $\hat{y}$ axis as a result of ferromagnetic resonance (FMR), induced by an external microwave driving field. The precessing magnetization injects spin angular momentum into the interfacial region of the 2DEG, generating a nonequilibrium spin accumulation polarized along $\hat{y}$. This spin accumulation arises from the balance between spin injection due to the magnetization dynamics and spin diffusion within the 2DEG. Through the inverse Edelstein effect, the interfacial spin accumulation is converted into a charge current flowing along the $\hat{x}$ direction in the 2DEG channel.} 
\label{Edelstein_cartoon}
\end{figure}
On the experimental side, spin–charge interconversion driven by the Edelstein effect has been intensively explored at oxide interfaces. In particular, LaAlO$_3$/SrTiO$_3$-based heterostructures have emerged as a versatile platform for investigating both DEE and IEE, owing to their gate-tunable carrier density, interfacial inversion symmetry breaking, and sizable Rashba interaction. Pioneering experiments demonstrated current-induced spin polarization and efficient spin-to-charge conversion at these oxide interfaces, establishing the relevance of the Edelstein mechanism beyond conventional semiconductor systems. 
In particular, a series of works~\cite{zpvy-t4d4, vicente2021spin, bibes2011ultrathin, trier2022oxide, massabeau2025inverse} provided key experimental evidence of the IEE in oxide heterostructures by means of spin pumping from an adjacent ferromagnet, revealing significant conversion efficiencies and their strong dependence on interfacial electronic reconstruction and spin–orbit coupling strength. 
These studies also highlighted the crucial role of orbital hybridization and multiband effects in determining the spin–charge conversion response~\cite{trama2022tunable,johansson2021spin}, setting oxide interfaces apart from simpler Rashba 2DEGs and undercovering their potential for oxide-based spin–orbitronic devices.
Despite the fact that the IEE is experimentally more accessible via a measurable charge current, the development of a simple theoretical description remains still challenging. This difficulty stems from the fact that, in the IEE, the magnetization is introduced as a localized perturbation at the boundary of the system, thereby breaking translational invariance along one spatial direction. This situation differs fundamentally from the DEE, where the driving electric field can be treated as a homogeneous external perturbation. Several theoretical approaches have been proposed to address this issue~\cite{shen2014microscopic, kato2004current, silov2004current, gorini2012onsager, sanchez2013spin, vignale2016theory, PhysRevResearch.3.033137, seibold2017theory}. Microscopic descriptions based on a generalized Buttiker formalisms~\cite{PhysRevB.46.12485,PhysRevLett.57.1761} have been employed for confined geometries such as Hall bars, using numerical calculations~\cite{zpvy-t4d4}. In parallel, more elementary semiclassical approaches based on the Boltzmann equation have also been developed, mainly in the context of topological insulators~\cite{Geng2017IEE}.
In this work, we study the IEE in a 2DEG confined along the $\hat{y}$ direction, in proximity to a ferromagnet, as depicted in Fig.~\ref{Edelstein_cartoon}. We formulate a theoretical description based on semiclassical Boltzmann approach with proper boundary conditions and this framework allows for the analytical evaluation of key physical observables, such as the induced charge current and the spin-torque. This central outcome of our work, not obtained before, provides direct physical insight and enable a transparent and quantitative benchmarking with experiments. This is particularly advantageous for complex oxide interfaces, such as LAO/STO, where material-specific parameters and interfacial inhomogeneities play a crucial role, and where analytical modeling offers a robust reference framework for interpreting and guiding spin–orbitronic measurements.
The paper is organized as follows. In Secs.~\ref{sec:model} and \ref{sec:approach} we introduce, respectively, the Rashba model and the semiclassical Boltzmann framework employed in our analysis. In Sec.~\ref{sec:charge current}, we find the analytical expression for the electric current by adopting an appropriate Boltzmann distribution function, and we discuss its behavior as a function of the chemical potential and the Rashba coupling strength. In Sec.~\ref{sec:spin current}, we study the spin-current and its related continuity equation in the context of Boltzmann approach. Finally in Sec.~\ref{sec:conclusions} we discuss our results and give the perspectives of our work.

\section{The two-dimensional Rashba gas}\label{sec:model}
%%%%%%%%%%%%%%%%
We consider the setup schematically shown in Fig.~\ref{Edelstein_cartoon}, which is designed to investigate the inverse Edelstein effect at the interface between a ferromagnet and a two-dimensional electron gas (2DEG) with Rashba spin–orbit coupling. In the physical situation of interest, the ferromagnet is driven into ferromagnetic resonance, giving rise to a time-dependent magnetization that injects spin angular momentum into the adjacent 2DEG. This spin pumping mechanism leads to a nonequilibrium spin accumulation in the interfacial region, whose steady-state value results from the dynamical balance between spin injection and spin diffusion.

In order to capture the time-independent consequences of this nonequilibrium spin accumulation, we adopt a simplified description in which the effect of the magnetization dynamics is represented by an effective static magnetization oriented along the $\hat{y}$ direction. This static exchange field should be understood as the temporal average of the precessing magnetization and provides an effective description of the stationary spin accumulation established in the 2DEG under dynamical equilibrium conditions. Importantly, this approximation does not imply that a static ferromagnet in proximity to a Rashba 2DEG would generate a stationary charge current. Rather, the time-independent treatment is intended to isolate the steady-state response associated with the presence of a spin accumulation maintained by spin pumping and spin relaxation processes.

\begin{figure} [t]
\centering
\includegraphics[width=0.45\textwidth]{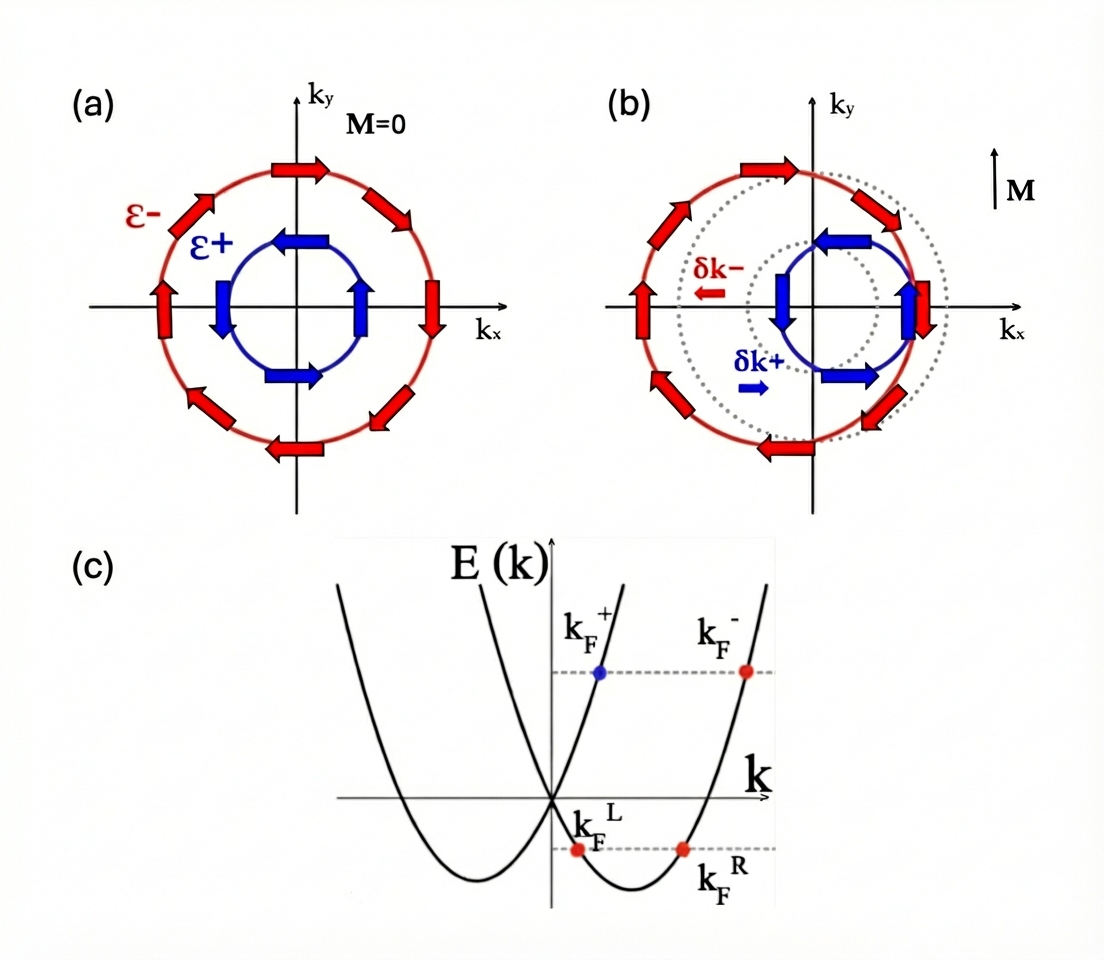}
 \caption{Schematic representation of Fermi surfaces in HDR (a) at the equilibrium and (b) with an applied magnetic field $\mathbf{M}= M_y \hat{y}$. If an external magnetization is applied, the Fermi surfaces are shifted by a $\delta k$ in opposite directions. (c) Energy dispersion of the Rashba 2DEG and Fermi momenta at fixed energy for high-density regime (HDR) and low-density regime (LDR). Blue and red dots indicate the branches of the energy dispersion, respectively $+$ and $-$.} 
 \label{Edelstein}
\end{figure}

Within this framework, the system can be modeled as a Rashba electron gas in contact with a ferromagnet described by an effective exchange Hamiltonian $\hat{H}_{\mathrm{FM}} = h_y \ \boldsymbol{\sigma}\cdot\hat{y}$, where $\boldsymbol{\sigma}$ denotes the Pauli matrices and $h_y$ is the effective exchange splitting. The Hamiltonian of the Rashba 2DEG acting on the spinor wave function then reads
%%%%%%%%%%%%%%%

\begin{equation}
\hat{H}=\frac{k^2}{2 m}+\alpha \hat{z}\cdot( \boldsymbol{\sigma} \times \textbf{k}),
\label{hamiltonianaiso}
\end{equation}
where the first term represents the kinetic energy, $\mathbf{k}$ is the momentum, $k$ its modulus, $m$ is the effective carrier mass, and we set $\hbar=1$ throughout. The second term corresponds to the RSOC contribution, where $\alpha$ is the coupling strength.
The resulting band structure is illustrated in Fig.~\ref{Edelstein}. 
The spectrum consists of two chiral bands whose dispersion is given by:
\begin{equation}
 \varepsilon_{\kvec}^{\nu}= \frac{k^2}{2m} +\nu k\alpha,
 \label{energy_spectrum}
\end{equation}
where $\nu=\pm$ is the chiral index. 
The occupation of the bands depends on the chemical potential $\mu$. For $\mu \ge 0$, over the band crossing, both chiral bands provide carriers that contribute to transport, whereas for $\mu < 0$ only one band is occupied~\cite{PhysRevLett.116.166602}. Therefore is convenient to distinguish between a high-density regime (HDR), defined by $\mu \ge 0$, and a low-density regime (LDR), defined by $\mu < 0$. This distinction is particularly relevant since transport properties are determined by the Fermi momenta $k_F$, whose expressions differ in the two regimes. In the HDR, both chiral bands contribute to transport, and the corresponding Fermi momenta read
\begin{equation}
 k_F^\nu= -\nu k_0 + \sqrt{k_0^2 + 2m\mu}, 
 \qquad \textrm{(HDR)}
\label{kHD}
\end{equation}
where $k_0=\alpha m$.
The two values of $k_F^\nu$ correspond to the blue and red dots shown in Fig.~\ref{Edelstein}(c).
In the LDR, only a single band is occupied, yet two distinct Fermi momenta are associated with it. In this case, one can distinguish the left ($k_F^L$) and right ($k_F^R$) Fermi momenta corresponding to the branch $\varepsilon^-_{\mathbf{k}}$, whose expressions read
\begin{equation}
 k_F^{\eta}= k_0 -\eta \sqrt{k_0^2 + 2m\mu}, 
 \qquad \textrm{(LDR)}
\label{kLD}
\end{equation}
which are both indicated by the red dots in Fig.~\ref{Edelstein} and where $\eta=\pm=R/L$ distinguishes between the left and right carriers for the lower band. 
The difference between these two regimes comes from the band chirality and the direction of the group velocity of the charge carriers. While the standard chiral operator, $S=\hat{n}\cdot(\boldsymbol{\sigma}\times \mathbf{k})/k$, successfully discriminates between the upper and lower Rashba bands, it does not provide an equivalent classification at the level of the carriers transport. Above the Rashba band-crossing point, the two Fermi surfaces possess opposite chiralities; however, below the crossing they are the same. For this reason it is essential to introduce a transport-chiral index, defined as $\eta = -\nu(\hat{\mathbf{v}} \cdot \hat{\mathbf{k}})$, with $\hat{\mathbf{v}}=\nabla_{\mathbf{k}}\varepsilon^\nu/|\nabla_{\mathbf{k}}\varepsilon^\nu|$ is the normalized group velocity, and $\hat{\mathbf{k}}=\mathbf{k}/k$. This is the same index that enters in Eq.~\eqref{kLD}.
In the following, we use benchmark parameters in agreement with experimental measurements on oxide interfaces~\cite{Ilani2018SrTiO3}, i.e., an effective mass $m= 0.7 m_e$, a RSOC coupling strength $\alpha= 0.006-0.01$~eV\AA, and a lattice parameter $a= 3.905$\AA. Considering the electron mean free path on a surface between oxides, the transport time is fixed at $\tau= 10^{-11} \ s$ ~\cite{Geng2017IEE}. In order to properly employ a semiclassical approach, it is necessary to work in the diffusive regime. Looking at Fig.~\ref{Edelstein_cartoon}, we therefore choose a length much larger than the electron mean free path, and accordingly set $L_x= 500$~$\mu$m.
In the following section, we introduce the semiclassical Boltzmann formalism to investigate the IEE.

\section{Semiclassical Boltzmann Approach} \label{sec:approach}

As shown in Fig.~\ref{Edelstein_cartoon}, a ferromagnet covering the region $x \le 0$ is placed in proximity to a Rashba 2DEG. The effect of the magnetization dynamics is described in terms of an effective static exchange field oriented along the $\hat{y}$ direction, representing the time-averaged magnetization under ferromagnetic resonance conditions. This effective field induces a steady nonequilibrium spin accumulation in the 2DEG, which is converted into a charge current via the inverse Edelstein effect. 

To study the spin and charge transport, we adopt the semiclassical Boltzmann formalism.
The inhomogeneous and stationary Boltzmann equation reads
\begin{equation}
 \textbf{v}_{\kvec} \cdot \frac{\partial f (\textbf{r},\textbf{k} )}{\partial \textbf{r}} = -\frac{f(\textbf{r},\textbf{k} ) - \left \langle f \right \rangle }{\tau},
\label{eq_bol_f}
\end{equation}
where the classical non-equilibrium distribution function $f(\textbf{r},\textbf{k} )$ of electrons in the system depends on the phase-space point $(\textbf{r},\textbf{k})$, with $\textbf{k}$ and $\textbf{r}$ being the momentum and coordinate of an electron, $\tau$ is the transport time and $\left \langle f \right \rangle$ represents the angular average of the distribution function. We thus assume that the scattering processes involved are elastic and do not introduce any preferred direction in momentum space; as a result, they are isotropic and drive the distribution function toward its angular average.
Since $f$ is translational invariant along $\hat{y}$, the left term of the Boltzmann equation can be rewritten as $v_{x\kvec} \frac{\partial f(x,\kvec) }{\partial x}$, with $v_{x\mathbf{k}}=\pdv{\varepsilon_{\mathbf{k}}}{k_x}$ that is the $x$-component of the velocity $\textbf{v}_{\kvec}$. 
We consider a regime of weak nonequilibrium and assume that the distribution function can be written in terms of a local deviation from equilibrium. Specifically, we adopt the ansatz
\begin{equation}
 f(x,\kvec)\approx f_0 + \bigg ( -\frac{\partial f_0}{\partial \varepsilon_{\kvec}} \bigg )g(x,v_{x\kvec}), 
\label{f_distribution}
\end{equation}
where $f_0$ is the equilibrium Fermi-Dirac distribution function and $g(x,v_{x\kvec})$ encodes the spatially dependent correction associated with a local shift of the chemical potential. Within this framework, the stationary Boltzmann equation determines the function $g$, describing the first correction due to the external magnetization. 
Here we make explicit the dependence of $g$ on the group velocity $v_{x\mathbf{k}}$, since it selects the carriers of the Rashba 2DEG that effectively contribute to scattering processes. We note that, at zero temperature, $-\pdv{f_0}{\varepsilon_{\mathbf{k}}}=\delta(\varepsilon_{\mathbf{k}}-\mu)$, which restricts the contribution to carriers at the Fermi energy, with fixed polar angle, $\theta=\arctan\!\left(\frac{k_y}{k_x}\right)$. Therefore in the following we will evaluate the group velocity at the Fermi momenta given in Eqs.~(\ref{kHD},~\ref{kLD}).
The equation for the distribution correction $g$ reads
\begin{equation}
 v_{x}^{\nu/\eta}\,\frac{\partial g(x,v_x^{\nu/\eta})}{\partial x}
 = - \frac{g(x,v_x^{\nu/\eta})- \langle g^{\nu/\eta}\rangle}{\tau},
 \label{eqforg}
\end{equation}
where $v^{\nu/\eta}_{x}= |v_F^{\nu/\eta}|\cos\phi$, with $\phi$ denoting the angle between the group velocity and the $\hat{x}$ direction, and
\begin{equation}
v^{\nu/\eta}_F
= \left. \pdv{\varepsilon^{\nu}_{\mathbf{k}}}{k} \right|_{k=k_F^{\nu/\eta}}
= \left(\frac{k_F^{\nu/\eta}}{m} + \nu\alpha\right).
\end{equation}
Finally, the angular average of the distribution function is then defined as
\begin{equation}
 \langle g^{\nu/\eta}\rangle
 = \frac{1}{2\pi} \int_0^{2\pi} d\phi \, g(x,v_x^{\nu/\eta}).
\end{equation}
The indices $\nu$ and $\eta$ are to be chosen according to the HDR and LDR, respectively, see Eqs.~\eqref{kHD} and \eqref{kLD}.

The problem is fully specified once appropriate boundary conditions are imposed, which determine the form of the function $g(x,v_{x\kvec}^\nu)$. In the limit $x\to 0^{-}$ the proximity of the ferromagnet induces a spin injection polarized along the $y$ direction into the 2DEG.
The presence of such a perturbation at $x\to 0^-$ induces an energy splitting $\delta\varepsilon^\nu=-\nu h_y \cos(\theta)$ on the eigenvalues of Hamiltonian~\eqref{hamiltonianaiso} where $\theta$ is the angle formed by the magnetization with the $\hat{y}$ direction. This can be conveniently considered as a boundary condition on $g(x,v_{x \kvec}>0)$ for the incoming carriers that is opposite in for the two bands. (see Appendix~\ref{app:g} for detailed derivation),
\begin{equation}
 g(x=0,v_{x\kvec}^{\nu}>0) = -\nu h_y \cos(\theta)\Theta(v_{x\kvec}^{\nu}).
 \label{condizione_cotorno}
\end{equation}
On the other hand, at the boundary $x=L_x$, electrons incoming from the right side of the system are in equilibrium; therefore, the corresponding boundary condition for the function $g$ reads:
\begin{equation}
 g(x=L_x,v_{x\kvec}^{\nu}<0) = 0.
\end{equation}
Integrating the first-order linear differential Eq.~\eqref{eqforg} and taking the boundary conditions for $g$ into account, it is possible to obtain a formal solution for the distribution function:
\begin{equation}
\begin{split}
 & g(x,v_{x\kvec}^{\nu})= \Theta(v_{x\kvec}^{\nu}) \Biggl \{ (\mp h_y \cos{\theta}) e^{-\frac{x}{v_{x\kvec}^{\nu} \tau}} + \\
 & \int\limits_0^x \langle g^{\nu} \rangle e^{-\frac{x-\xi}{v_{x\kvec}^{\nu}\tau}} \frac{1}{v_{x \kvec}^{\nu} \tau} d\xi \Biggr\} + \\
 & \Theta(-v_{x \kvec}^{\nu}) \left\{ \int\limits_{L_x}^x \langle g^{\nu} \rangle e^{-\frac{x-\xi}{v_{x \kvec}^{\nu}\tau}} \frac{1}{v_{x \kvec}^{\nu} \tau} d\xi \right\}.
\end{split}
\label{autocons}
\end{equation}
In Refs.~\cite{gaiardoni2025edelstein, Geng2017IEE} an approximate self-consistent solution of Eq.~(\ref{autocons}) is provided by a linear approximation of $\langle g^\nu\rangle$. Although the solution for the function 
$g$ is given by an integral and self-consistent expression, it can only be solved numerically taking into account the boundary conditions mentioned above. In our work, instead, we choose to use a different expression for the function $g$. In fact, it is straightforward to verify that, for a Rashba 2DEG, a monopole expansion of $g$ provides a solution of Eq.~(\ref{autocons}), as a consequence of the isotropic nature of the model. 
Owing to the strongly anisotropic boundary conditions, however, this solution is valid only in the region where the distribution function is isotropic, $\lambda < x < L-\lambda$, with $\lambda = |v_F|\tau$ the mean free path. We note that $\lambda$ is the only intrinsic length scale of the problem and therefore sets the spatial range over which this solution applies.
Within this range of validity, the approach with the linearized $\langle g^\nu\rangle$ and the monopole expansion are equivalent; however, the latter allows for a fully analytical treatment, providing additional insight into the structure of the distribution function.
Since in our geometry $v_x \propto \cos(\phi)$, the main anisotropic correction is captured by retaining only the isotropic component and the first angular harmonic, whereas higher harmonics, generated by the boundary conditions, are naturally suppressed over distances of order $\lambda$.
Thus, an analytic solution can be obtained by means of the monopole expansion:
\begin{equation}
g^{\nu/\eta}= g_0^{\nu/\eta} + g_1^{\nu/\eta} \cos(\phi),
\label{espressione_g}
\end{equation}
where we introduced the notation $g^{\nu/\eta} \equiv g(x,v_x^{\nu/\eta})$ and $\nu/\eta$ are used respectively in HDR or LDR. 
\\Substituting the expression of $g^{\nu/\eta}$ in Eq.~\eqref{eqforg} and considering the boundary conditions, we can find the coefficient $g_0^{\nu/\eta}$ and $g_1^{\nu/\eta}$ (see Appendix \ref{monopol_solution} for the details): 
\begin{equation}
\begin{split}
 g_0^{\nu/\eta}&=- \frac{c^{\nu/\eta}}{ v_F^{\nu/\eta} \tau} x + c_0^{\nu/\eta}\\
 g_1^{\nu/\eta}&=c^{\nu/\eta},
\end{split}
\label{g0_g1}
\end{equation}
where 
\begin{equation}
\begin{split}
 c^{\nu/\eta}&= \frac{2B^{\nu/\eta} v_F^{\nu/\eta} \tau}{4v_F^{\nu/\eta}\tau + \pi L_x}\\
 c_0^{\nu/\eta}& = B^{\nu/\eta}\left(\frac{2}{\pi} - \frac{4v_F^{\nu/\eta}\tau}{4v_F^{\nu/\eta}\tau + \pi L_x }\right)
\end{split}
\end{equation}
with $B^{\nu}=-\nu h_y$ for the HDR, while $B^{\eta}=-\eta h_y$ for the LDR.
The use of the monopole expansion of the function $g$ makes it possible to analytically calculate the charge current that flows in the system due to the IEE.\\

\section{Electric current due to inverse Edelstein effect} \label{sec:charge current}

Within the semiclassical Boltzmann approach, the electric current generated by IEE can be written as:
\begin{equation}
 I_c= I_0 \sum\limits_{\nu/\eta} \int v_{x}^{\nu/\eta} g^{\nu/\eta} \biggl ( -\frac{\partial f_0}{\partial \varepsilon_{\kvec}^{\nu}} \biggl )k \, dk\, d \theta,
\label{corrente_IEE_formula}
\end{equation}
where we set $I_0 = \frac{e L_y}{4\pi^2m a^3}$, while we use dimensionless variables in the integral. 
As discussed in Sec.~\ref{sec:approach}, the contributions of $g$ and $\langle \hat{v}_x \rangle_{\mathbf{k}}^\nu$ are already evaluated at the Fermi momenta selected by $-\pdv{f_0}{\varepsilon_{\mathbf{k}}}$.\\
\noindent Using Eq.~(\ref{espressione_g}) for the $g(x,v_{x}^{\nu/\eta})$ we can find an analytical expression for the electric current, which differs for HDR and LDR and read:
\begin{equation}
\begin{split}
 I_c(\mu\ge0)= & I_0 \pi \sum_\nu g_1^\nu k_F^\nu= \\
 &\frac{4 h_y m \pi \alpha \sqrt{m (2 \mu + m \alpha^2)} \tau}{L_x m \pi + 4 \sqrt{m (2 \mu + m \alpha^2)} \tau},
\label{formula_corrente_HDR}
\end{split}
\end{equation}
and 
\begin{equation}
\begin{split}
 &I_c(\mu<0)= I_0 \pi \sum_\eta g_1^\eta k_F^\eta= \\
 & \frac{4 h_y m \pi (2 \mu + m \alpha^2) \tau}{L_x m \pi + 4 \sqrt{m (2 \mu + m \alpha^2)} \tau}.
\label{formula_corrente_LDR}
\end{split}
\end{equation}
Let us note that the current is independent of the spatial coordinate $x$ and therefore is conserved along the $x$ direction. This follows from the fact that the analytic expression depends only on the $g_1$ coefficient given in Eq.~\eqref{espressione_g} and is consistent with the isotropization requirement for the equilibrium distribution function in Eq.~\eqref{eq_bol_f}. 
\begin{figure}[t!]
 \includegraphics[width=0.45\textwidth]{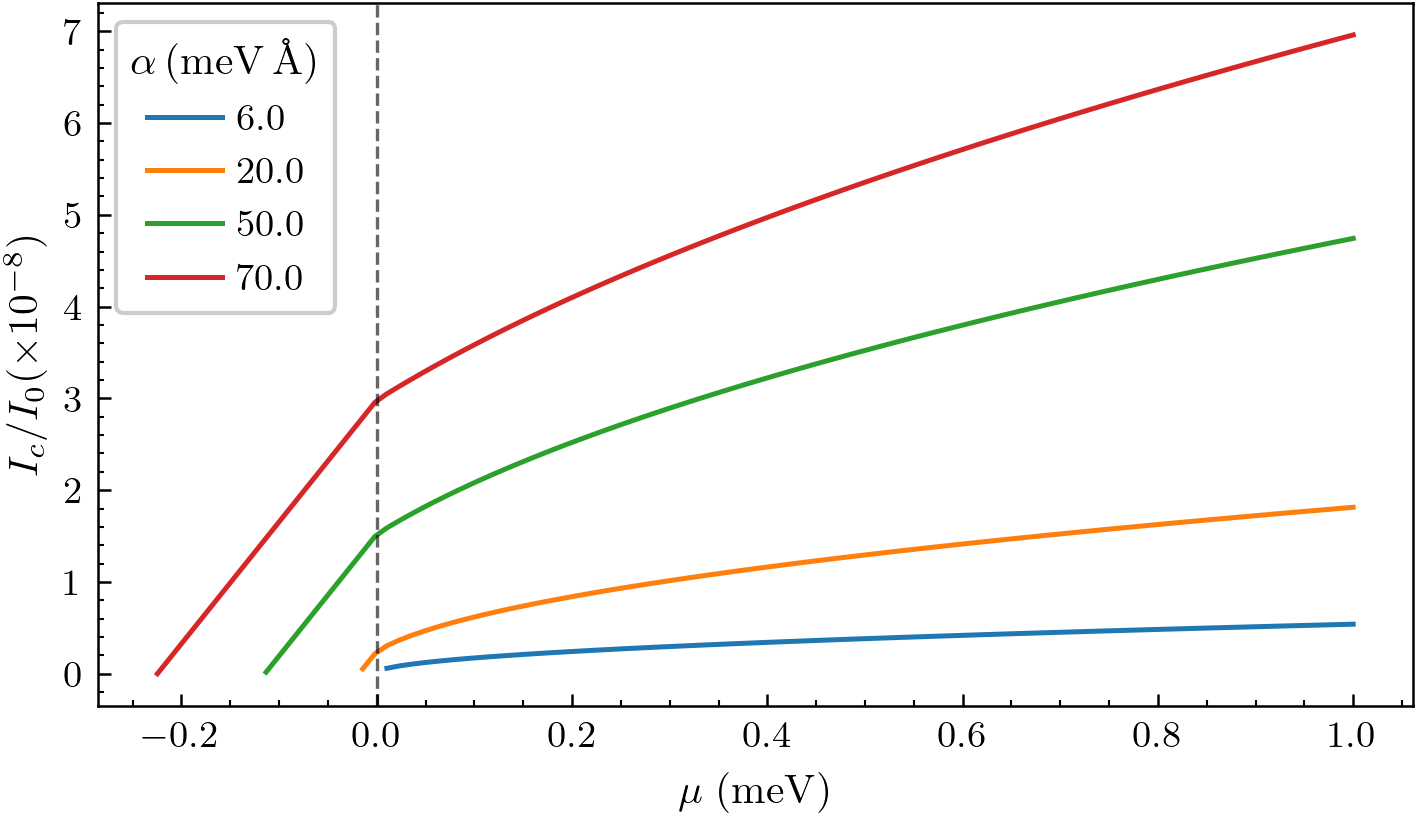}
 \caption{Electric current as a function of the chemical potential $\mu$ for different values of the Rashba parameter $\alpha$, where $I_0= 5.63$~A. The dashed line at $\mu=0$ represents the crossover between LDR and HRD.}
 \label{CORRENTEELETTRICA_IEE_1}
 \end{figure}
 %\hspace{0.01\textwidth}
 % Pannello b
\begin{figure}[t!]
\includegraphics[width=0.45\textwidth]{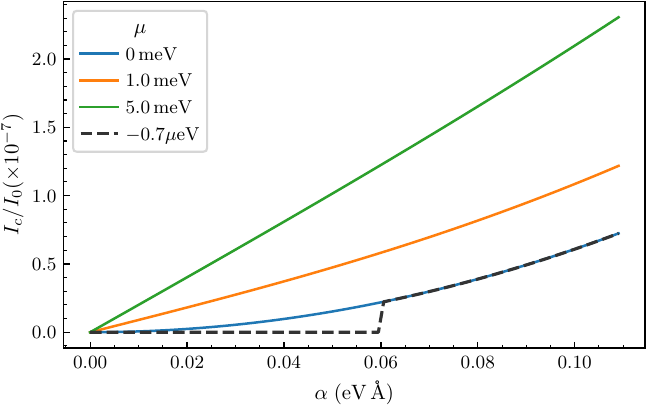}
 \caption{Electric current as a function of the Rashba parameter $\alpha$ for different values of the chemical potential. Here $I_0= 5.63$~A.}
 \label{CORRENTEELETTRICA_IEE_2}
\end{figure} 
In Fig.~\ref{CORRENTEELETTRICA_IEE_1} we show the electric current within the whole region of chemical potential and for different benchmark parameters of the Rashba coupling $\alpha$. For our parameter choice, $I_c\sim\mu$ in LDR, since $L_x m\gg4\sqrt{m(2\mu+m\alpha^2)}\tau$. This differs from the HDR behavior which goes as $I_c\sim\sqrt{\mu}$. For $\mu<-m\alpha^2/2$, the expressions lose of validity.
In Fig.~\ref{CORRENTEELETTRICA_IEE_2} we show the electric current as a function of $\alpha$ for fixed values of chemical potential. When $\mu \gg m\alpha^2/2$, the current increases linearly with $\alpha$. Within the HDR, for $\mu < m\alpha^2/2$, the predicted behavior becomes quadratic in $\alpha$, as indicated by Eq.~\eqref{formula_corrente_HDR}. The LDR, on the other hand, is characterized by the range $-m\alpha^2/2 < \mu < 0$, where the current also exhibits an approximately quadratic dependence on $\alpha$, see Eq.~\eqref{formula_corrente_LDR}. 
As expected, when $\alpha=0$, i.e., in the absence of RSOC, spin-to-charge conversion cannot occur and the current vanishes. Also this limit in captured by Eq.~\eqref{formula_corrente_HDR}. 

\section{Spin current and spin continuity equation}\label{sec:spin current}
Having access to the distribution function gives us the opportunity to study also the spin-response to the system, which is indeed expected due to the shift of the Fermi surfaces. In systems with spin–orbit coupling, the spin density is not a conserved quantity, and therefore a consistent definition of spin current is problematic~\cite{PhysRevLett.96.076604} and requires some caution. 
As widely done in literature, we introduce the spin-current operator, defined as the symmetrized product of the velocity and spin operators, $j^j_i=\{v_i, {\sigma}_j\}/2$, and then calculate its average~\cite{PhysRevB.73.113305}. This operator describes the rate at which the spin component $j$ is transported along the spatial direction $i$ by the motion of the electronic wave packet. The symmetrization ensures hermiticity and accounts for the fact that, in the presence of spin–orbit interactions, $v_i$ and $\sigma_j$ do not generally commute. Therefore, the definition of the spin current reads
\begin{equation}
 I_S= I_{S_0}\sum\limits_{\nu} \int \frac{1}{2}\langle\{v_{x}, \sigma_y\}\rangle^{\nu/\eta} \,g^{\nu/\eta} \biggl (- \frac{\partial f_0}{\partial \varepsilon_{\kvec}^{\nu}} \biggl )k \, dk d \theta
\label{expression_Is}
\end{equation}
where $I_{S_0}=\frac{I_0}{2e}$ and we choose all quantities in the integral as dimensionless.  Interestingly, the mean value of the anticommutator is analytically given by:
\begin{equation}
 \frac{1}{2}\langle\{v_{x}, \sigma_y\} \rangle^{\nu/\eta}= -\nu \frac{k_F^{\nu/\eta}}{m} \cos^2(\theta) - \alpha,
 \label{eq:anticommutator}
\end{equation}
for which we can identify two contributions: one depending on the carrier momentum and the chirality of the band, and a constant term coming from the presence of RSOC.
We can explicitly calculate the spin current expression and get in the HDR:
\begin{equation}
\begin{split}
&I_S(\mu>0)= \\&I_{S_0}\sum_\nu \bigg \{ \left(-\nu \frac{\pi}{m} k_F^\nu - 2\pi \alpha\right)g_0^\nu \frac{k_F^\nu}{|v_F^\nu|} \bigg \}=\\
&\frac{
 8 h_y \mu \left(
 m \pi (L_x - x)
 + 2 \sqrt{m \left( 2 \mu + m \alpha^2 \right)} \, \tau
 \right)
}{
 L_x \pi \sqrt{m \left( 2 \mu + m \alpha^2 \right)}
 + 8 \mu \tau
 + 4 m \alpha^2 \tau
}
\end{split}
\label{spin_current_HDR}
\end{equation}

and in LDR:
\begin{equation}
\begin{split}
 &I_S(\mu<0)=\\
 &I_{S_0}\sum_\eta \bigg \{ \eta^2 \left(\frac{\pi}{m} k_F^\eta - 2\pi \alpha\right)g_0^\eta \frac{k_F^\eta}{|v_F^\eta|} \bigg \}=0.
\end{split}
\label{spin_current_LDR}
\end{equation}

\begin{figure}[t!]
 % Pannello a
 \includegraphics[width=0.45\textwidth]{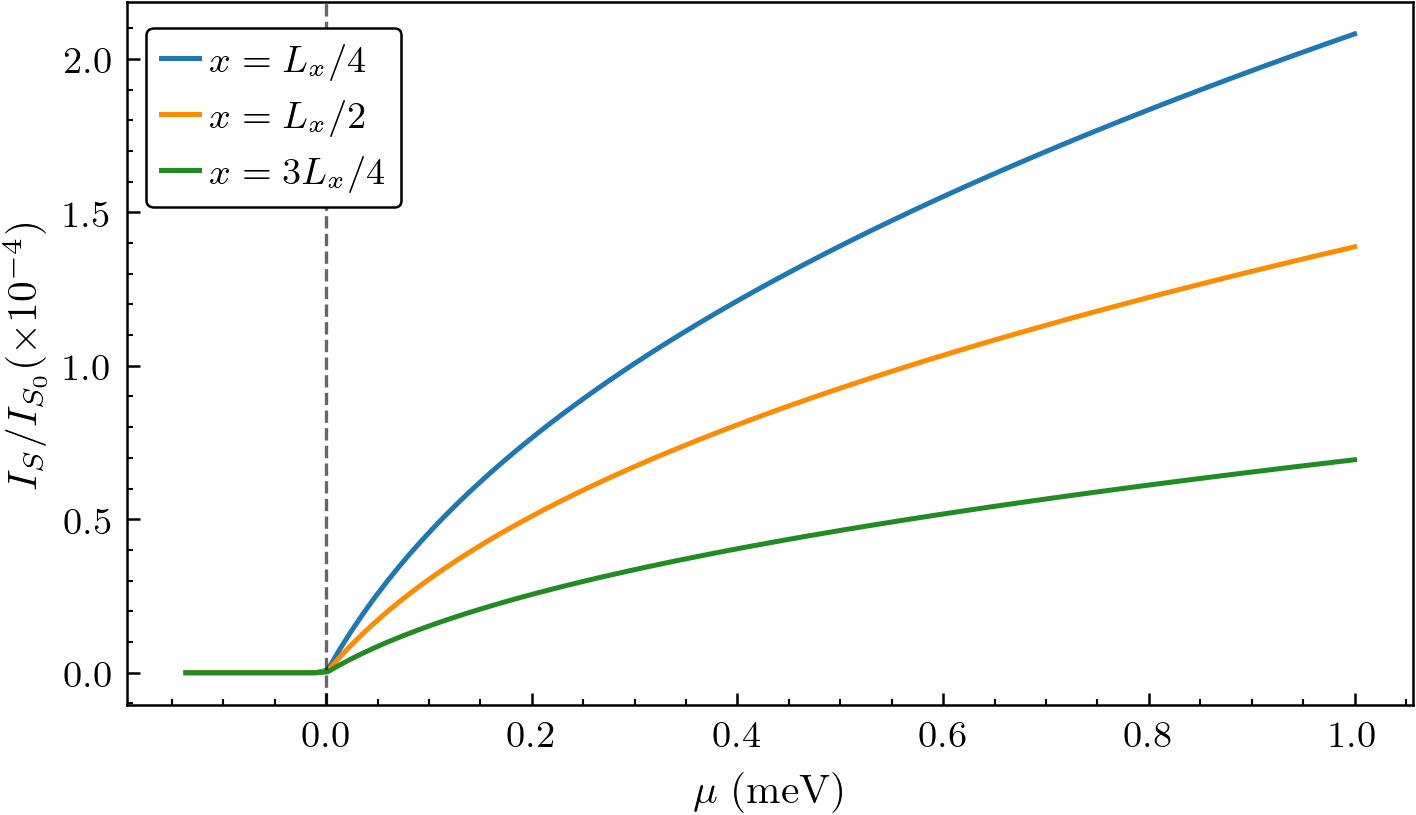}
 \caption{Spin current as a function of the chemical potential for different values of $x$, for a fixed value of Rashba parameter $\alpha= 54$~meV\AA, considering the electronic correlation between the spin and the group velocity.}
 \label{CORRENTESPIN_IEE_correlazioni}
\end{figure}
In Fig.~\ref{CORRENTESPIN_IEE_correlazioni} we show the spin current for benchmark values of $x$. Contrarily to the electric currents defined in Eqs.~(\ref{CORRENTEELETTRICA_IEE_1},~\ref{CORRENTEELETTRICA_IEE_2}), the spin current is not conserved due to the spin precession along the $x$-direction. This is clearly reflected in the spin current expression  (~\ref{spin_current_HDR}) that depends on $x$ via the coefficient $g_0$ (see Eq.~\eqref{g0_g1}).
In Fig.~\ref{CORRENTESPIN_IEE_correlazioni}, the spin current is zero below the band crossing and grows only for chemical potentials above it. In particular, Eq.~\eqref{spin_current_LDR}, vanishes due to the presence of the factor $\alpha$ in Eq.~\eqref{eq:anticommutator} which does not depend on the transport-chirality index of the carrier. Moreover, it is interesting to note that a vanishing spin current is a characteristic feature of the LDR, i.e. when the Fermi surfaces do not experience a relative shift with respect to each other but are displaced in the same direction in momentum space.

Since the IEE gives rise to a spin-polarized current along the $\hat{y}$ direction, which decays along the $\hat{x}$ direction of the system, it is natural to ask whether torque-like phenomena can emerge in such a system. So, to derive the expression for this quantity, it is convenient to start from the continuity equation for the spin $\partial S_i/\partial t = -\nabla\cdot \mathbf{J}_i + T_i$ ~\cite{PhysRevB.81.045307}, where $S_i$ is the spin density along $i=x,y,z$, $\mathbf{J}_i$ is the spin current density carrying the $i$-th spin component and $T_i$ is the torque density.
To derive such an equation in the diffusive regime, we may directly use the expression for the magnetization:
\begin{equation}
\begin{split}
 S_i= \sum_{\mathbf{k}, \nu} \langle \sigma_i \rangle_{\mathbf{k}}^{ \nu} f^\nu(x,\mathbf{k}).
\end{split}
\label{magn}
\end{equation}
By taking the time derivative of Eq.~\eqref{magn} and considering the Boltzmann equation
\begin{equation}
 \pdv{f}{t} = - v_x \pdv{f}{x} - \frac{\delta f}{\tau}
\end{equation}
we obtain
\begin{equation}
\begin{split}
\frac{\partial S_i}{\partial t}
= \sum_{\nu/\eta} \int &d^2k\,
\langle \sigma_i \rangle^{\nu/\eta}
\left( - \frac{\partial f_0}{\partial \varepsilon_{\kvec}^{\nu}} \right) \cdot\\ &\bigg[
- v_x^{\nu/\eta} \frac{\partial g^{\nu/\eta}}{\partial x}
- \frac{1}{\tau} g^{\nu/\eta}
\bigg],
\label{magn2}
\end{split}
\end{equation}
\noindent where we already considered the quantities evaluated at the Fermi momenta selected by $\delta(\varepsilon_\kvec^\nu-\mu)$. We can identify the second term in Eq.~(\ref{magn2}) as the torque. Let us note that since we are considering a static magnetization along the $y$-axis, the only component of the torque that is different from zero is along $\hat{y}$ and its expression is: 
\begin{equation}
\begin{split}
 T_y= &-T_0 \sum_\nu \int d^2k \,\langle \sigma_y \rangle^{\nu/\eta} v_{x}^{\nu/ \eta}\frac{\partial g^{\nu/\eta} }{\partial x}\biggl (- \frac{\partial f_0}{\partial \varepsilon_{\kvec}^{\nu}} \biggl )= \\
 &-4 T_0 \pi h_y m\frac{\sqrt{m(2\mu+\alpha^2 m)}}{4\tau \sqrt{m(2\mu+\alpha^2 m)}+ \pi m L_x }
 \end{split},
\end{equation}
in HDR and
\begin{equation}
\begin{split}
 T_y= 
 &-\frac{4 T_0 \pi h_y m^2 \alpha}{4\tau \sqrt{m(2\mu+\alpha^2 m)}+ \pi m L_x }
 \end{split}
\end{equation}
in LDR, where $T_0= 1/ma^4$.
The fact that $T_y$ is the only torque component indicates that, rather than a torque, we observe a modulation of the spin current along the $x$-direction in the system. This effect is commonly known as a spin sink.

\noindent By reconsidering the spin continuity equation, Eq.~\eqref{magn2}, we find that the first term can be identified with the divergence of the expectation value of the spin–current density operator defined in Eq.~\eqref{expression_Is} only when $v_x$ commutes with $\sigma_y$ and both commute with the Hamiltonian $\hat{H}$. Upon integration along the transverse direction, this term reduces to
\begin{equation}
 I_S= I_{S_0}\sum_{\nu/\eta} \int d^2k \langle \sigma_y \rangle^{\nu/\eta} v_{x}^{\nu/\eta} g^{\nu/\eta} \biggl (- \frac{\partial f_0}{\partial \varepsilon_{\kvec}^{\nu}} \biggl ).
\label{corrente_spin}
\end{equation}
where $I_{S_0}$ accounts for this integration performed along the $y$-direction.
In Appendix~\ref{spincurrent_eqcontinuity}, we show that this condition holds when the Rashba spin–orbit coupling becomes progressively less relevant, i.e. for higher values of the chemical potential $\mu$.

\section{Conclusions}\label{sec:conclusions}

In this work, we investigate spin–charge conversion in a two-dimensional electron gas with Rashba spin–orbit coupling, focusing on the inverse Edelstein effect. The system was described by the Rashba Hamiltonian, which captures the essential consequences of inversion-symmetry breaking at oxide interfaces, such as LaAlO$_3$/SrTiO$_3$ and LaAlO$_3$/KTaO$_3$. 

Nonequilibrium transport was treated within the semiclassical Boltzmann formalism, with boundary conditions accounting for spin injection from an adjacent ferromagnet and relaxation processes within the two-dimensional electron gas. 
Within the diffusive regime and the relaxation time approximation, the inhomogeneous Boltzmann problem admits a fully analytical treatment: by retaining the first angular harmonic in the angular expansion of the non-equilibrium distribution function, we derive closed-form expressions for the inverse Edelstein effects, the charge current (and the corresponding spin-current response) in both the high- and low-density regimes. These analytical formulas provide benchmark estimates of spin–charge conversion as a function of chemical potential, Rashba coupling, scattering time, and device length.
The induced electric current exhibits a strong dependence on the chemical potential, with a marked qualitative change when the chemical potential crosses the band-touching point, as shown in Fig.~\ref{CORRENTEELETTRICA_IEE_1}. 
As the chemical potential increases, the current initially grows linearly and displays an abrupt change upon crossing the band-touching point. This behavior reflects the discontinuity in the density of states when transitioning between the low- and high-density regimes.

The spin current, which is not conserved in the presence of spin–orbit coupling, was evaluated starting from its operator definition and reformulated within the Boltzmann framework. We find that the spin current vanishes below the band-crossing point and becomes finite only for positive chemical potentials. This result can be traced back to the nature of the inverse Edelstein effect: at low carrier densities and in the presence of a finite magnetization, the Fermi contours undergo a rigid shift without generating a relative displacement between the Rashba-split bands.
We also highlight the limitations of the Boltzmann approach in this context. In particular, a formulation of the spin-current continuity equation within semiclassical Boltzmann theory, including spin-torque contributions, is reliable only in the regime of weak Rashba spin--orbit coupling or when the chemical potential lies sufficiently far from the band-crossing point.
A more general description for low-energy regime would likely require a theory based on the density matrix
capable of properly capturing spin coherence effects ~\cite{Schwab_2011, PhysRevB.55.5908}. 
Overall, our results show how spin–orbit coupling, band filling, and interface-induced boundary conditions jointly determine the efficiency of spin–charge conversion in Rashba systems. These findings provide a transparent framework for interpreting inverse Edelstein measurements through oxides interfaces. Extensions of this work could include a self-consistent treatment of the Boltzmann equation and the incorporation of magnetization dynamics, which would allow access to nonlinear and time-dependent spin–orbitronic effects.

\section{Acknowledgments}
We acknowledge R. Capecelatro for useful discussions. This work received support from the PNRR MUR Project No. PE0000023-NQSTI (TOPQIN and SPUNTO)  and from the project INNOVATOR (National centre HPC, big data and quantum computing). M.T. acknowledges PNRR MUR Project No. PE0000023-NQSTI (SOC-OX CUP E63C22002180006). F.R. acknowledges funding from Ministero dell’Istruzione, dell’Università e della Ricerca (MIUR) for the PRIN project STIMO (Grant No. PRIN 2022TWZ9NR). This work was financed by Horizon Europe EIC Pathfinder under the Grant IQARO No. 101115190. 

\appendix
\section{Determination of the distribution function variation} \label{app:g}
To justify the choice of our boundary conditions in Eq.~\eqref{condizione_cotorno}, as discussed in the main text, we start by writing the Hamiltonian of the Rashba 2DEG at the interface between the ferromagnet
\begin{equation}
    \hat{H}=\frac{k^2}{2 m}+\alpha \hat{z}\cdot( \boldsymbol{\sigma} \times \textbf{k})+h_y\sigma_y.
\end{equation}
In this condition, the general distribution function $f_{\kvec}$ in the Boltzmann equation can be expanded for small values of $h_y$:
\begin{equation}
 f_{\kvec}
 \approx f_0 + \delta f_{\kvec} = f_0 + \bigg ( \frac{\partial f_{\kvec}}{\partial \varepsilon_{\kvec}^{\pm}} \frac{\partial \varepsilon_{\kvec}^{\pm}}{\partial h_y} \bigg ) \bigg |_{h_y=0} h_y,
\end{equation}
where 
\begin{equation}
    \varepsilon_{\kvec}^{\nu}= \frac{k_x^2 + k_y^2}{2m} +\nu \sqrt{h_y^2 - 2h_y \alpha k_x + (k_x+ k_y)^2\alpha^2}
    \end{equation}
is the energy spectrum. Performing the derivative of the energy respect to the parameter $h_y$ and by substituting $k_x=k\cos{\theta}$ and $k_y=k\sin{\theta}$, $\delta f_{\kvec}$
takes the form:
\begin{equation}
 \delta f_{\kvec} = \mp \bigg( -\frac{\partial f_0}{\partial \varepsilon_{\kvec}^{\pm}} \bigg ) h_y \cos{\theta},
\label{deltaf}
\end{equation}
where $\kvec=k \ (\cos \theta, \sin \theta)$.
This result is consistent with our assumption for the boundary condition at $x=0$ of Eq.~\eqref{condizione_cotorno}.

\section{Analytical calculation of the monopole solution of the function $g^{\nu/\eta}$}
\label{monopol_solution}
In this appendix, we explicitly derive the coefficients for the monopole expansion of the function $g^{\nu/\eta}= g_0^{\nu/\eta} + g_1^{\nu/\eta} \cos(\phi)$. We begin by computing the function $\langle g^{\nu/\eta} \rangle$:
\begin{equation}
 \langle g^{\nu/\eta} \rangle= \frac{1}{2\pi} \int\limits_0^{2\pi} d\phi \left( g_0^{\nu/\eta} + g_1^{\nu/\eta}\cos{\phi}\right) = g_0^{\nu/\eta}.
\end{equation}
Substituting the expressions for $g(x, v^{\nu/\eta}_{x})$ and $\langle g^{\nu/\eta} \rangle$ into Eq.~\eqref{eqforg}, we obtain:
\begin{equation}
 v^{\nu/\eta}_x \frac{\partial }{ \partial x} \left(g_0^{\nu/\eta}(x)+ g_1^{\nu/\eta}(x)\cos{\phi}\right)= - \frac{g_1^{\nu/\eta}(x) \cos{\phi}}{\tau}.
 \label{eq:eq_for_monopole}
\end{equation}
Multiplying by $\cos{\phi}$ and integrating the left- and right-hand sides of Eq.~\eqref{eq:eq_for_monopole} with respect to $\phi$, we obtain
\begin{equation}
 |v_F| \frac{\partial g_0^{\nu/\eta}}{\partial x} = -\frac{g_1^{\nu/\eta}}{\tau} \implies g_0^{\nu/\eta}= -\frac{g_1^{\nu/\eta}}{|v_F| \tau} x + c^{\nu/\eta}_0.
\end{equation}
A second equation can be obtained by integrating the equation directly with respect to $\phi$, so we can write:
\begin{equation}
\frac{\partial g_1^{\nu/\eta}}{\partial x}=0 \implies g_1^{\nu/\eta}= c^{\nu/\eta}
\end{equation}
So, we can see that $g_1^{\nu/\eta}$ does not depend on $x$.
To impose our boundary conditions, we must distinguish the calculations performed in the HDR and LDR, due to the different group velocity directions of the carriers.
Indeed, the group velocity and momentum have the same orientation on both Fermi surfaces in HDR, whereas in LDR the group velocity of the inner Fermi surface is opposite in sign to its momentum. Therefore, particular care must be taken when evaluating integrals involving a change of variables between $\phi$, the angle in velocity space, and $\theta$, the angle in momentum space. 
In the HDR, $\theta=\phi$ for both carriers, so we can write: 
\begin{equation}
\begin{split}
 \langle g(x=0,v^{\nu}_x>0)\rangle= &\frac{1}{2\pi} \int\limits_{-\pi/2}^{\pi/2} d\phi\, (-\nu h_y \cos(\phi))=\\
 &-\frac{\nu h_y}{\pi}=\frac{B^{\nu}}{\pi} \\
 \langle g(x=L_x,v^{\nu}_x<0)\rangle= 0
\end{split}
\end{equation}
Instead, in the LDR, $\phi= \theta + \pi$ for $\eta=1$, while $\phi=\pi$ for $\eta=-1$. Therefore, we obtain
\begin{equation}
\begin{split}
 \langle g(x=0,v^\eta_x>0)\rangle= &\frac{1}{2\pi} \int\limits_{-\pi/2}^{\pi/2} d\phi\, (-\eta h_y \cos{\phi})=\\
 &-\frac{\eta\, h_y}{\pi}= \frac{B^{\eta}}{\pi}\\
\langle g(x=L_x,v^{\eta}_x<0)\rangle= 0.
\end{split}
\end{equation}

We can then set the expressions derived from the boundary conditions equal to the original definition of $g(x, v^{\nu/\eta}_{x})$, thereby obtaining, for example, in the HDR:
\begin{equation}
\begin{split}
 &\frac{1}{2\pi}\int\limits_{-\pi/2}^{\pi/2} d\phi \,\left(g^{\nu}_0 + g^{\nu}_1 \cos{\phi}\right)=\frac{-\nu h_y}{\pi} \implies \\
 & \implies\pi c_0^{\nu} + 2 c^{\nu}= -2\nu h_y \\
& \frac{1}{2\pi}\int\limits_{\pi/2}^{3\pi/2} d\phi \,\left(g^{\nu}_0 + g^{\nu}_1 \cos{\phi}\right) = 0 \implies \\
&\implies\left( - \frac{L_x \pi}{|v_F^\nu| \tau} -2 \right ) c^{\nu} + c^{\nu}_0 \pi =0 
\end{split}
\end{equation}

and we can do the same in the LDR.
Solving the system of two equations yields the coefficients $c^{\nu / \eta}$ and $c_0^{\nu / \eta}$ in the two regimes:
\begin{equation}
\begin{split}
 &c^{\nu}=\frac{2 B^{\nu} |v_F^\nu| \tau}{4 |v_F^\nu| \tau + \pi L_x}\\
 & c^{\nu}_0 = B^{\nu} \left( \frac{2}{\pi} - \frac{4 |v_F^\nu| \tau}{4 |v_F^\nu| \pi\tau + \pi^2 L_x} \right)
\end{split}
\end{equation}
in the HDR, and 
\begin{equation}
\begin{split}
 &c^{\eta}=\frac{2 B^{\eta} |v_F| \tau}{4 |v_F^\eta| \tau + \pi L_x}\\
 & c^{\eta}_0 = B^{\eta} \left( \frac{2}{\pi} -\frac{4 |v_F^\eta| \tau}{4 |v_F^\eta| \pi\tau + \pi^2 L_x} \right)
\end{split}
\end{equation}
in the LDR.

\section[\appendixname~\thesection]{The evaluation of the spin current from the continuity equation}
\label{spincurrent_eqcontinuity}
\begin{figure}[t]
 \includegraphics[width=0.45\textwidth]{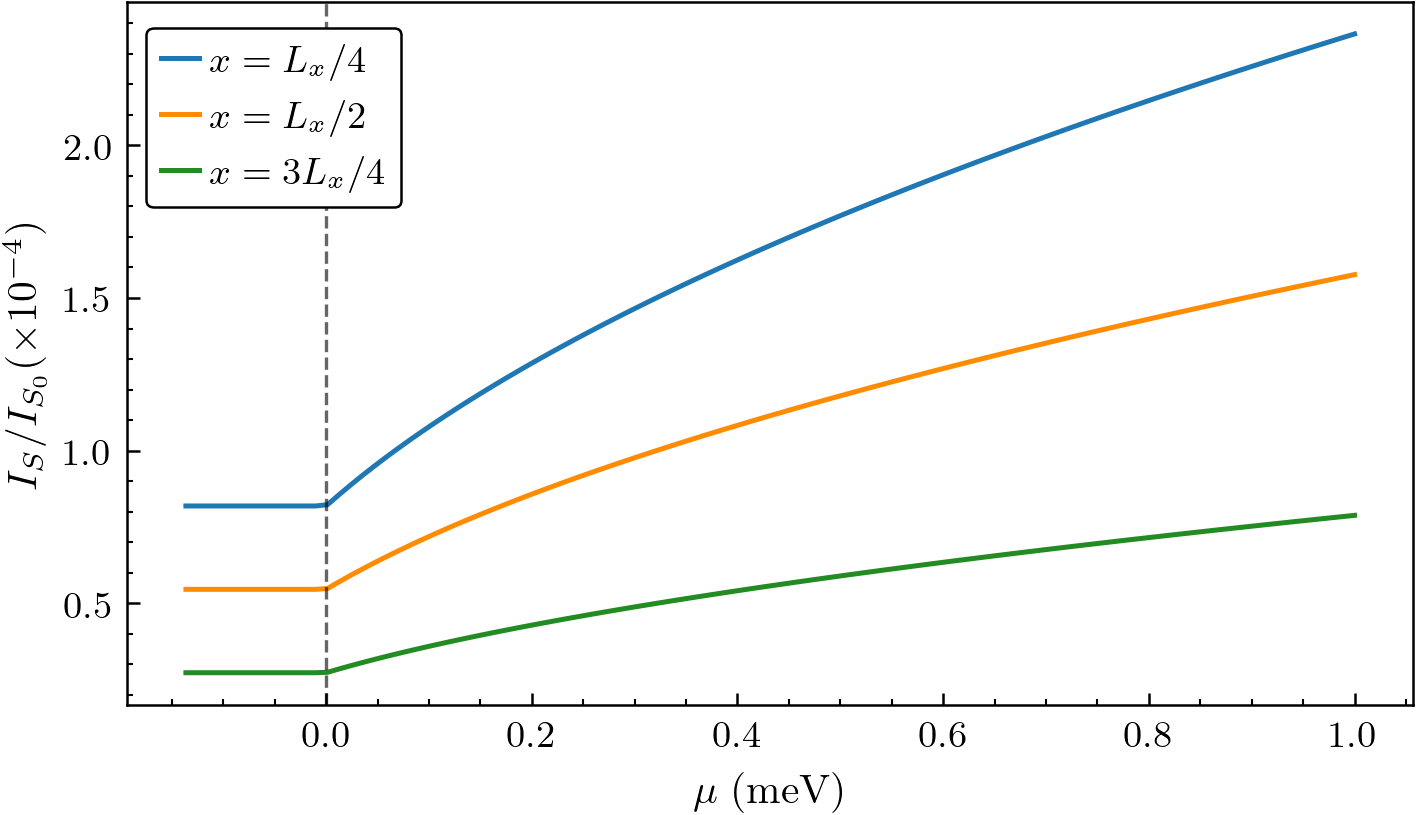}
 \caption{Normalized spin current as a function of the chemical potential for different values of $x$ and for a fixed value of $\alpha=54$~meV\AA.}
 \label{CORRENTESPIN_IEE}
\end{figure}

In this appendix, starting from the definition of the spin current in Eq.~\eqref{expression_Is}, we show that, when reduced to the single-band limit or to weak spin–orbit coupling, it recovers Eq.~\eqref{corrente_spin}, namely, the spin-current expression coming from the continuity equation coincides with the expectation value of the spin-current operator.
It is straightforward to demonstrate the identity
\begin{equation}
    \frac{1}{2}\langle\{v_{x}, \sigma_y\} \rangle^{\nu/\eta}= v_x^{\nu/\eta} \langle\sigma_y\rangle^{\nu/\eta} -\alpha\sin^2(\theta).
\end{equation}
Therefore if we consider a system in which the RSOC is sufficiently weak, we may also consider the group velocity and the spin as decoupled and mutually independent and Eqs.~\eqref{expression_Is} and \eqref{corrente_spin} coincide.
However, since the Fermi momenta at which these expressions are evaluated depend on the chemical potential, within the energy window considered near the band crossing, the RSOC is far from negligible. Indeed, as we can see in Fig.~\ref{CORRENTESPIN_IEE}, which shows the spin current as a function of the chemical potential, a clearly nonphysical result emerges: at the band bottom—where the group velocity vanishes, as discussed above—the spin current does not go to zero.
This behavior suggests that writing the continuity equation directly in terms of the magnetization constitutes an approximation that breaks down in Rashba systems at low energies, calling for a more refined treatment based on kinetic theory or on a density-matrix formulation.
\vspace{0.03cm}

\begin{figure}[H]
 \includegraphics[width=0.45\textwidth]{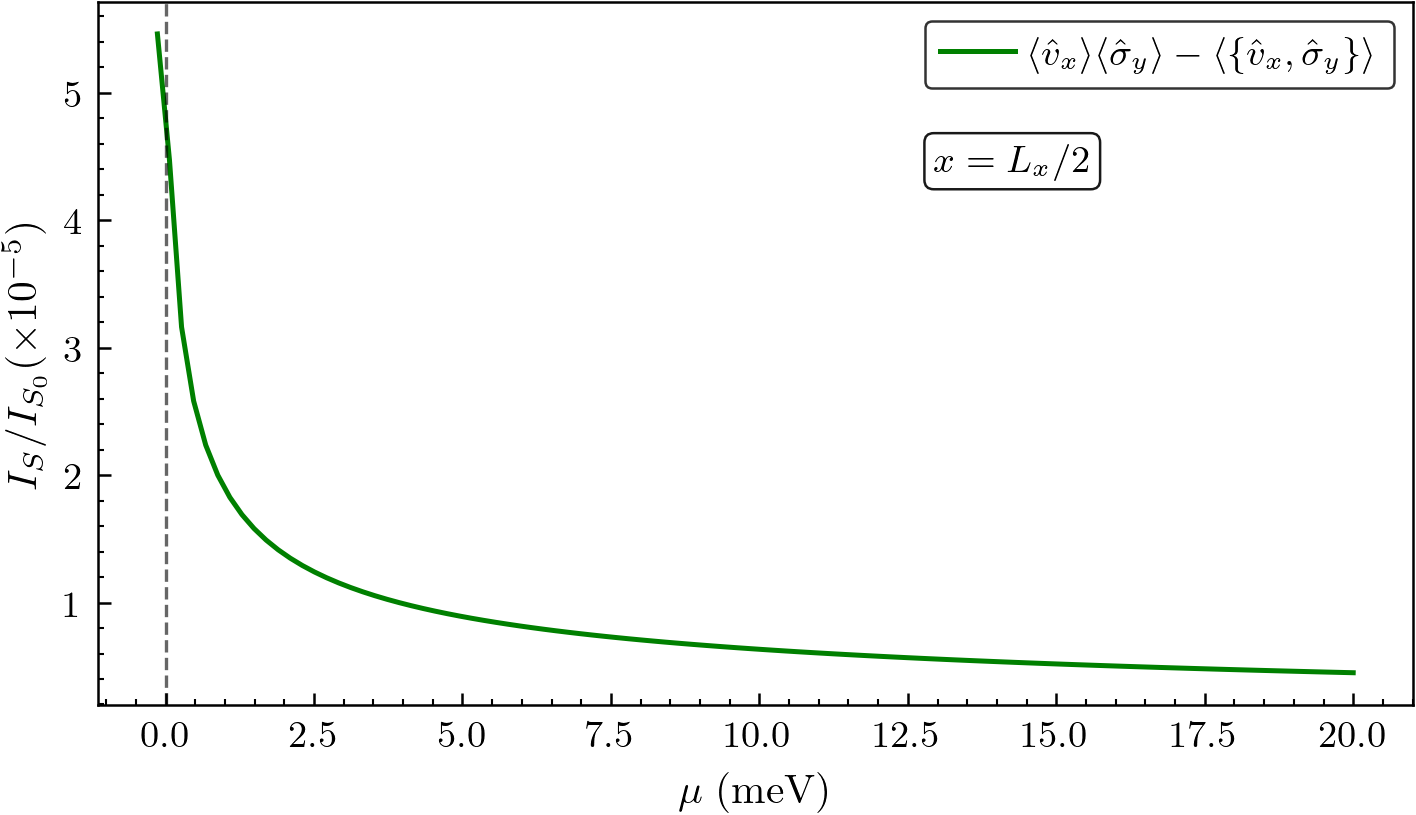}
 \caption{Difference of the spin current computed using Eqs.~\eqref{spin_current_HDR}-\eqref{spin_current_LDR} and Eq.~\eqref{corrente_spin} as a function of the chemical potential for $\alpha=54$~meV\AA~and for $x=L_x/2$. 
 }
 \label{CORRENTESPIN_IEE_confronto}
\end{figure}

However, as shown in Fig.~\ref{CORRENTESPIN_IEE_confronto}, for large values of the chemical potential (i.e., when $k_F\gg k_0$), the correct expression of Eq.~\eqref{expression_Is} accounting the contribution of the SOC and the pathological one, expression~\eqref{corrente_spin}, derived from the continuity equation, gradually converges. In fact, as expected, as the chemical potential increases, the Rashba spin–orbit coupling becomes progressively less relevant and the two expressions effectively coincide.

\bibstyle{unsrt}
\bibliography{Bibliography.bib}
\vspace{12pt}
\color{red}
\end{document}